\documentclass{amsart}
\usepackage{blkarray}
\usepackage{cite}
\usepackage{stmaryrd}
\usepackage{mma}

\usepackage{amssymb,graphicx,enumerate,color}

\renewcommand{\P}{\mathcal{P}}

\usepackage{graphicx}
\usepackage[colorlinks=true,citecolor=black,linkcolor=black,urlcolor=blue]{hyperref}

\numberwithin{equation}{section}
 \theoremstyle{plain}
\newtheorem{theorem}{Theorem}[section]

\newtheorem{definition}[theorem]{Definition}

\theoremstyle{remark}

\usepackage[framemethod=tikz]{mdframed}

\begin{document}

\makeatletter
\def\imod#1{\allowbreak\mkern10mu({\operator@font mod}\,\,#1)}
\makeatother

\author{Jakob Ablinger}
   \address{Research Institute for Symbolic Computation, Johannes Kepler University Linz, Altenbergerstrasse 69, 4040 Linz, Austria}
   \email{jakob.ablinger@risc.jku.at}

\author{Ali Kemal Uncu}
   \address{Austrian Academy of Sciences, Johann Radon Institute for Computational and Applied Mathematics, Altenbergerstrasse 69, 4040 Linz, Austria}
   \email{akuncu@ricam.oeaw.ac.at}

\title[\texttt{\lowercase{q}F\lowercase{unctions}} package for $q$-series and partitions]{\texttt{\lowercase{q}F\lowercase{unctions}} - A Mathematica package for $q$-series and partition theory applications} 

\begin{abstract} We describe the qFunctions Mathematica package for $q$-series and partition theory applications. This package includes both experimental and symbolic tools. The experimental set of elements includes guessers for $q$-shift equations and recurrences for given $q$-series and fitting/finding explicit expressions for sequences of polynomials. This package can symbolically handle formal manipulations on $q$-differential, $q$-shift equations and recurrences, such as switching between these forms, finding the greatest common divisor of recurrences, and formal substitutions. Here, we also extend the classical method of the weighted words approach. Moreover, qFunctions has implementations that automate the recurrence system creation of the weighted words approach as well as a scheme on cylindric partitions.
\end{abstract}
   


\thanks{Research of the first author is supported by the Austrian Science Fund FWF grant SFB F50 (F5009-N15). Research of the second author is supported by the Austrian Science Fund FWF, SFB50-07, SFB50-09 and SFB50-11 Projects.}

\date{\today}
   \maketitle

\section{Introduction}

The theory of partitions and its accompanying field $q$-series enjoy a large and ever increasing community. Problems of interest, backgrounds of researchers, and methods used in this population show great variety. One common point of it all is that the explicit computations in $q$-series tend to be time-consuming, dreary, and error-prone. Therefore, the necessity of any and all computer algebra assistance is clear. Andrews~\cite{Andrews_Comps,Andrews_q_series} has been encouraging and guiding the introduction of computer algebra systems to this field by sharing his intuitions for decades. To this end, some toolsets have been implemented to help researchers semi-automatically carry out their calculations. One such example worth mentioning is the~\texttt{qseries} Maple package by Garvan~\cite{qSeries}. As the needs of the community become better defined, more involved software implementations got created especially by the Research Institute for Symbolic Computation Combinatorics group. As a relevant example, implementations by Koutschan, Paule, Riese, Schneider, etc. \cite{HolonomicFunctions, Sigma, qMultiSum, qZeil} are just some of computer algebra implementations that can automatically find and prove recurrence relations of $q$-series with hypergeometric terms, which is a highly common theme in $q$-series.

Although the common themes such as solving recurrences will always be relevant, new approaches keep on being introduced. As an example recently, experimental research gained great momentum. Especially the methods and computer implementations by Kanade--Russell~\cite{Kanade_Russell, IdentityFinder, Mathew_Thesis} for finding conjectural identities attracted a lot of interest. These conjectural identities and related results already led to numerous discoveries~\cite{Kagan1,Kagan2,Kagan3,BerkovichUncu7,BerkovichUncu8,BerkovichUncu9,Bringmann_JS_Mahlburg,Uncu2,Uncu3} of both combinatorial and analytic nature. Simultaneously, independent of the mentioned experimental methods, the method of weighted words introduced by Alladi--Gordon \cite{Generalized_Schur} got revisited by Dousse and others~\cite{Dousse_Unif, Bringmann_Dousse, Dousse_Lovejoy, Dousse_Siladic, Dousse_revisit, Dousse_Primc}. Lastly, a new promising scheme involving cylindric partitions for finding identities has been introduced by Corteel, Foda, and Welsh~\cite{Corteel, Foda}.

All things considered, it became clear that the community can benefit from a computer implementation that concentrates on these techniques that have not yet been included in a software repository before. The \texttt{qFunctions}\footnote{The package \texttt{qFunctions} downloaded at \url{http://www.risc.jku.at/research/combinat/software/qFunctions}.} Mathematica package provides tools to systematically guess $q$-difference equations and recurrences in the spirit of Bringmann--Jennings-Shaffer--Mahlburg's proofs of some of the Kanade--Russell conjectures, symbolically handles the recurrence creation of the method of weighted words that is in the spirit of Dousse's recent work, automates the $q$-difference system creation of the method introduced by Corteel--Welsh, and do experimental polynomial fitting with known hypergeometric objects in the spirit of the \texttt{RRtools} Maple package of Sills. This new package has both symbolic and experimental tools built in, and it is designed (and encouraged) to be used together with other symbolic computation implementations such as \texttt{HolonomicFunctions} and \texttt{Sigma}.

The organization of this paper is as follows. Section~\ref{Section_Def} has the necessary definitions and some background information on $q$-differential/shift equations, and recurrences. Section~\ref{Section_q_diff} builds up on the $q$-shift equations/recurrences and explains the guessing functionality of the \texttt{qFunctions} package. In Section~\ref{Section_WW}, the method of weighted words and its implementation is presented. At the end of the section, the weighted words method is extended to handle a larger set of partitions and this extension is also explained. Corteel and Welsh's recent scheme on cylindric partitions and related functionality of the package are introduced in Section~\ref{Section_Cylindric}. Section~\ref{Section_q_fit} has some usage notes on experimentally trying to fit a sequence of $q$-polynomials as a sum of some $q$-hypergeometric objects. Moreover, each section includes relevant examples that exemplify the use of the package.

With that all laid out, we start by loading the \texttt{qFunctions} package into Mathematica:
\begin{mma}
\In \mathbf{< \hspace{-1.5mm}< } |qFunctions.m|\\
\Print HolonomicFunctions package is required to use the full functionality of this package.\\
\Print \vspace{-.2cm}\\
\Print \fbox{qFunctions by Jakob Ablinger and Ali K. Uncu -- $\copyright$ RISC Linz -- Version 1.0 (\today)}\\
\end{mma}

\section{Definitions and Background}\label{Section_Def}
We follow the classical definitions of partitions and related objects \cite{Theory_of_Partitions, Gasper_Rahman}. A finite list of non-negative integers is called a \textit{composition}. A \textit{partition} is a composition, where all parts are positive and in an non-increasing order. The sum of all the parts of a composition or a partition $\pi$ is called the \textit{size} of said object, and denoted with~$|\pi|$. When the counts of these objects are in question, the attention shifts to generating functions and this takes the study of these combinatorial objects to partition theory's analytic counter part $q$-series. 

Let $q$, $a$, $a_i$'s and $b$ be variables, where $q$ is transcendental over $\mathbb{Q}$. Let $n$ and $m$ be non-negative integers. We first define the $q$-bracket
\begin{align} \nonumber [n]_q &= \frac{1-q^n }{ 1-q}.
\intertext{Then the $q$-factorial is}
\nonumber [n]_q! :&= \prod_{i=1}^{n} [i]_q.
\intertext{We define the $q$-Pochhammer symbol}
\nonumber (a;q)_n :&= (1-a)(1-aq)(1-aq^2)\dots(1-aq^{n-1}),\\ 
\nonumber (a;q)_\infty :&=\lim_{n\rightarrow \infty} (a;q)_n,
\intertext{and} 
\nonumber (a_1,a_2,\dots,a_k;q)_n :&= (a_1;q)_n(a_2;q)_n\dots(a_k;q)_n.
\intertext{The $q$-binomial coefficients are defined as} 
\nonumber {n+m\brack n}_q :&= \left\{\begin{array}{ll}\displaystyle \frac{(q;q)_{n+m}}{(q;q)_n(q;q)_m}, & n+m\geq m\geq0,\\[-1.5ex]\\ 0, &\text{otherwise}. \end{array} \right. 
\intertext{Now we introduce the $q$-trinomial coefficients, which was first defined by Andrews--Baxter \cite{Andrews_Baxter}, as a $q$-analog of the ordinary trinomial coefficients,}
\label{Round_Tri_Def}
\left(\hspace{-2mm}\begin{array}{c}L,\, b\\ a \end{array};q\right)_2 &:= \sum_{n\geq 0 } q^{n(n+b)} \frac{(q;q)_L}{(q;q)_n(q;q)_{n+a}(q;q)_{L-2n-a}.}
\intertext{Then we define some useful variants of the $q$-trinomial coefficients in the spirit of Warnaar \cite{Warnaar}:}
\label{T_n_Def} T_n(L,a;q) &:= q^{(L-a)(L+a-n)/2} \left(\begin{array}{c}L,\, a-n\\ a \end{array};\frac{1}{q}\right)_2,\\
\label{t_n_Def} t_n(L,a;q) &:= q^{n(L-a)/2} \left(\begin{array}{c}L,\, a-n\\ a \end{array};q\right)_2.
\intertext{The function $t_0$ appears as $\tau_0$ in Andrews--Baxter's definition. Finally, we define two useful combinations}
\label{U_n_Def} U_n(L,a;q) &:= T_n(L,a+1;q) + T_n (L,a;q),\\
\intertext{and}
\label{V_n_Def} V_n(L,a;q) &:= T_{n+1}(L,a+1;q) + q^{(L-a)/2} T_{n}(L,a;q).
\end{align}
Note that Andrews--Baxter defined $T_n(L,a;q)$, $t_n(L,a;q)$ and $U(L,a;q)$ for $n=0,$ and $1$ in \cite{Andrews_Baxter} slightly differently than their definition here. The connection is a simple dilation $q\mapsto q^2$ in~\eqref{T_n_Def},~ \eqref{t_n_Def} for $n=0$ and 1, and \eqref{U_n_Def} for $n=0$. Also the function $V(L,a;q)$ defined by Sills \cite{Sills_FinRR,Sills_RRtools} is $V_0(L-1,a-1,q^2)$.

Let \begin{equation}\label{F_def}F(x)= \sum_{n\geq 0} a_n(q) x^n\end{equation} be a formal power series, where $a_n(q)$ are rational functions in $q$ with rational coefficients. The $q$-derivative is defined as \[\left( \frac{d}{dx}\right)_q F(x) := \frac{F(x) -F(xq)}{x(1-q)} = \sum_{n\geq 0} a_n(q) \frac{1-q^n}{1-q} x^{n-1}.\] A (linear) $q$-differential equation that a formal series $F(x)$ satisfies is equivalent to a $q$-recurrence for its coefficients $a_n(q)$. Another equivalent form is a (linear) $q$-shift equation ($q$-difference equation), where one openly writes the definition of the $q$-derivative and simplifies the equation. In general, It is possible to switch between these equivalent forms formally. We will only be focusing on linear $q$-difference and shift equations (and consequently only focus on linear $q$-recurrences).

One simple example is as follows. Assume that $F(x)$ satisfies the $q$-differential equation  \[q x^2 F''(x)- \frac{x (1+ q + q x)}{(1-q)} F'(x)+ \frac{1}{(1-q)^2}F(x) = 0. \]
We can write the definition of the $q$-derivatives in and simplify this $q$-differential equation to its equivalent $q$-shift equation form. After writing the definitions of the $q$-derivatives explicitly, we get
\[q x^2 \frac{F(x q^2) -(1-q) F(x q) + q F(x)}{q x^2 ( 1-q)^2}- \frac{x (1+ q + q x)}{(1-q)} \frac{F(x) -F(xq)}{x(1-q)}+ \frac{1}{(1-q)^2}F(x) = 0.\]
This simplifies to
\begin{equation}\label{F_q_shift}F(x q^2) + x q F(x q) - x q F(x) =0. \end{equation}
Now by using the definition of $F(x)$ and directly collecting the coefficients of $x^n$, we get the (linear) $q$-recurrence relation for the series coefficients of $F(x)$.
\begin{equation}\label{F_q_shift_open}
\sum_{n\geq 0} a_n(q) q^{2n} x^n + x q \sum_{n\geq 0} a_n(q) q^n x^n - x q \sum_{n\geq 0} a_n(q) x^n = 0,
\end{equation}
shows the relation
\begin{equation}\label{RR_rec} a_{n+1}(q) q^{2n+2} - q(1-q^{n+1}) a_{n}(q) = 0,\end{equation} for all $n\geq0$. Note that in equation \eqref{F_q_shift_open} the coefficient of the $x^0$ term sits outside, as an initial condition, of the recurrence \eqref{RR_rec}.

Reverting back from $q$-recurrences to $q$-shift equations and such can also be done in the same manner. If one mutliplies both sides of \eqref{RR_rec} with $x^n$ and sum over $n \geq 0$, we get \[\frac{1}{x}\sum_{n\geq 1} a_{n}(q) q^{2n} x^{n} - q \sum_{n\geq 0} a_n(q) x^n + q \sum_{n\geq 0} a_n(q) q^n x^n  = 0,\] after a shift in variable in the first sum. This sum can be related with the formal power series $F(x)$ (as defined in \eqref{F_def}) provided that we add and subtract the first term of the series, which we will represent with $\langle 1\rangle F(x)\,(:= a_0(q)\, )$. Then we get \[\frac{1}{x}\sum_{n\geq 0} a_{n}(q) q^{2n} x^{n} - q \sum_{n\geq 0} a_n(q) x^n + q \sum_{n\geq 0} a_n(q) q^n x^n - \frac{1}{x} \langle 1\rangle F(x) = 0.\] This is equivalent to \[F(x q^2) + x q F(x q) - x q F(x) + \langle 1\rangle F(x) =0\] which is slightly different than \eqref{F_q_shift}. This observation is equivalent to the note after \eqref{RR_rec}. Similar consideration needs to be made going to the $q$-differential equations. Switching between the $q$-shift/differential equations and $q$-recurrences one needs to be wary of the initial conditions. 

All these formal form switches can be done with the provided functions in the \texttt{qFunctions} package. The list of functions to this extend are \texttt{qDEToqSE, qDEToqRE, qSEToqRE, }etc. The syntax is demonstrated in the following example, where we start from a $q$-shift equation. All the recurrences and $q$-shift equations in \texttt{qFunctions} are always assumed to be equal to 0 and this equality is not written to keep the outputs and subsequent inputs clean.

\begin{mma}
\In |qse| = F[x q^2] - x^2 q F[x]; \\
\In |qSEToqDE|[|qse|, F[x]]\\
\Out \left(1-q x^2\right)  F[x]+(q-1) (q+1) x F'[x]+(q-1)^2 q x^2F''[x]\\
\In |qre| = |qSEToqRE|[|qse|, F[x], a[n]] \\
\Out a[n+2] q^{2 n+4}-q a[n] \\
\In |qREToqSE|[|qre|,a[n], F[x]]\\
\Out q^2 (-x) \langle x\rangle [F[x]]-\langle 1\rangle [F[x]]+F[q^2   x]-q x^2 F[x]\\
\In |qREToqDE|[|qre|,a[n], F[x]] \\
\Out -F[0]+q^2 (-x) F'[0]+\left(1-q x^2\right)  F[x]+(q-1) (q+1) x F'[x]+(q-1)^2 q x^2F''[x]\\
\end{mma}

In general, we will be using $\langle x^n\rangle F(x)$ to indicate the coefficient of the $x^n$ term of the formal power series $F(x)$ when we are dealing with the $q$-shift equations. Similarly, we use $F^{(n)} [0]$ to be the constant coefficient of the  $n$-th  $q$-derivative of $F(x)$ when we are working with $q$-differential equations.

Other useful and important tools such as looking for the greatest common divisor or making a substitution in $q$-recurrences and $q$-shift equations are also included in the package. The substitutions will be demonstrated in the next chapter, but here we would like to demonstrate listing values of a $q$-recurrence using initial values and finding a the greatest common divisor of two recurrences. 

\begin{mma}
\In |qREToList|[|qre|, a[n], \{-2,\{1,q\}\},5]\\
\Out \left\{1,q,q,1,\frac{1}{q^2}\right\}\\
\In |qREGCD|[|qre|, a[n+4] q^{4n+10} - a[n], a[n]]\\
\Out a[n+2] q^{2 n+3}- a[n]\\
\end{mma}

In the syntax of the \texttt{qREToList}, one gives the starting index and the initial conditions from that index on. In the example, the $\{-2,\{1,q\} \}$ portion of the function call defines the initial conditions as $a(-2)=1$ and $a(-1)=q$. In the same example we are asking for the list function to list 5 values of this sequence. In the output list, the first 2 values are the initial values that are given to the function and the following three values are the values of $a(0),$ $a(1),$ and $a(2)$, respectively. One can choose to include more initial values than the minimum necessary amount (the order of the recurrence) and in this case this function uses the last order many values of the list to calculate more initial values.

There is an independent predecessors of the \texttt{qFunctions} package, which has the same functionality explained in this section. The \texttt{qGeneratingFunctions} package of Kauers--Koutschan \cite{qGeneratingFunctions} can also do the mentioned formal manipulations, making lists from recurrence and initial conditions. One small caveat of this package is its insistence of calculating initial values formally at each step. This slows down the calculations especially when the orders are high. A user can choose to check initial conditions manually (by the listing option mentioned above) at his/her discretion when using the \texttt{qFunctions} package. 

One thing that \texttt{qFunctions} does not include that the \texttt{qGeneratingFunctions} package has is the closure properties for recurrences. One can automatically find the recurrence for the addition, multiplication, etc. of given sequences with known recurrences by employing the \texttt{qGeneratingFunctions} package.

\section{Guessing q-Shift Equations and \texorpdfstring{$q$}{q}-Recurrences}\label{Section_q_diff}

Given a formal power series $F(x)$ as defined in \eqref{F_def}, one can suspect the existence of a recurrence that the coefficients $a_n(q)$ satisfy (or equivalently a $q$-shift equation that $F(x)$ satisfies). The \texttt{qFunctions} package includes a customizable function to guess such a relation within the given boundaries. This is done by forming and solving a linear system, up to an automatically decided (or manually chosen) threshold.

For example, Let \[RR(x) = \sum_{n\geq 0} \frac{x^{n^2} x^n}{(q;q)_n},\] which is the well known generating function that is the analytic counterpart of the first Rogers--Ramanujan identities' gap conditions, where the exponent of $x$ counts the number of parts in the counted partitions \cite{Theory_of_Partitions}. We can easily guess a $q$-shift equation for $RR(x)$ using a list of the coefficients of $z^i$. 

If we expect to find a $q$-shift equation of order $2$ (meaning, the $q$-shift equation involves only $RR(x),$ $RR(xq)$ and $RR(xq^2)$), the degree of the coefficients in the $q$-shift equation is $\leq 3$ using the summands up to $n=30$, we write the following. 

\begin{mma}
\In |RRdata| = |CoefficientList|[ |Sum|[x^n q^{n^2} / |qPochhammer|[q, q, n], \{n, 0, 30\} ], x ];\\
\In |GuessqShiftEquation|[ |RRdata| , |RR|[x], 2,3]\\
\Print More restrictions/data needed. Example: set ExpansionOrder higher.\\
\Print C[i]'s are free constants.\\
\Out (-q^2 x^2 - q x^2 C[1] - x^2 C[2] - q^2 x C[3] - q x C[4] - x C[5] - 
    q^2 C[6] - q C[7] - C[8]) RR[
   x] + (q^2 x^2 + q x^2 C[1] + x^2 C[2] + q^2 x C[3] + q x C[4] + 
    x C[5] + q^2 C[6] + q C[7] + C[8]) RR[
   q x] + (q^3 x^3 + q^2 x^3 C[1] + q x^3 C[2] + q^3 x^2 C[3] + 
    q^2 x^2 C[4] + q x^2 C[5] + q^3 x C[6] + q^2 x C[7] + 
    q x C[8]) RR[q^2 x]\\
\end{mma}

Here, in the messy outcome, $C[i]$'s are free constants that one can specialize to get different guesses for the $q$-shift equations. As shown above, if the solution space of the linear system is multi-dimensional, a message is printed about changing the restrictions. This message is a suggestion to direct the user towards narrowing their search down to a single guess. There are many choices to increase restrictions. One can drop the order and/or degree in the guesses, use a larger data set, increase the series expansion order, assume a particular shape for the highest or the lowest ordered term, etc. We exemplify two options here:

\begin{mma}
\In |GuessqShiftEquation|[ |RRdata|, RR[x], 2, 1]\\
\Out -RR[x] + RR[q x] + q x RR[q^2 x]\\
\In |GuessqShiftEquation|[ |RRdata|, RR[x], 2, 3, |AddHighestOrderFactor| \shortrightarrow (1 - q ^2 x^2)]\\
\Out (-1 + q x) (1 + q x) RR[x] - (-1 + q x) (1 + q x) RR[q x] - 
 q x (-1 + q x) (1 + q x) RR[q^2 x]\\
\end{mma}

Note that in both cases, with the different restrictions, we were able to get a single guess. Also observe that these guesses are equivalent. In general, we find having the \texttt{AddHighestOrderFactor} and \texttt{FixLowestOrderCoefficient} useful in search for an identity that would be suitable for substitutions. 

Another modifier that one can use in guessing is the \texttt{StartingPoint}. If one believes that the initial terms (coefficients of $x^0, x^1,\dots, x^m$ for some $m$) might have some noise, they can start from a higher starting point (without modifying their data array). Moreover, so far all the guesses yielded and searched for homogeneous $q$-shift equations. By default the \texttt{qFunctions} package searches for homogeneous equations. One can change this and ask to include inhomogeneous $q$-shift equations in the search by employing the option \texttt{InhomogeneousEquation}$\shortrightarrow$\texttt{True}. 

Guessing recurrences for sequences is done in the same fashion, and we note that there are other recurrence guessers available that a user can choose from such as \texttt{Guess} \cite{Guess}, \texttt{RATE} \cite{RATE}, and \texttt{qGeneratingFunctions} \cite{qGeneratingFunctions}. Furthermore, once guessed, these recurrences can be proven/disproven using the symbolic computation implementations from the RISC Combinatorics group. These implementations include but not limited to the Mathematica packages \texttt{HolonomicFunctions} \cite{HolonomicFunctions}, \texttt{Sigma} \cite{Sigma}, \texttt{qMultiSum} \cite{qMultiSum}, and \texttt{qZeil} \cite{qZeil}. 

This method of experimentally guessing $q$-shift equations was recently employed by Bringmann--Jennings-Shaffer--Mahlburg \cite{Bringmann_JS_Mahlburg} as the backbone of their proofs for 7 of the Kanade--Russell conjectures \cite{Kanade_Russell, IdentityFinder, Mathew_Thesis}. We would like to follow the relevant steps of one of their proofs as a longer example here. For that we start by one of the proven Kanade--Russell conjectures \cite[Theorem 1.1]{Bringmann_JS_Mahlburg}:

\begin{theorem}[Bringmann, Jennings-Shaffer, Mahlburg, 2019]
\begin{equation}
\label{H1_1} \sum_{i,j,k\geq 0} (-1)^k \frac{q^{(i+2j+3k)(i+2j+3k-1)+3k^2 + i+6j+6k}}{(q;q)_i(q^4;q^4)_j(q^6;q^6)_k} = \frac{1}{(q,q^4,q^6,q^8,q^{11};q^{12})_\infty}
\end{equation}
\end{theorem}

In their proof, Bringmann--Jennings-Shaffer--Mahlburg start with the sum side of this sum-product identity and define \[H_1(x) := \sum_{i,j,k\geq 0} (-1)^k \frac{q^{(i+2j+3k)(i+2j+3k-1)+3k^2 + i+6j+6k}}{(q;q)_i(q^4;q^4)_j(q^6;q^6)_k} x^{i+2j+3k}.\] They then guess a $q$-shift equation over the exponent of $x$. They reduce the $q$-shift equations and recurrences by using substitutions till they find a recurrence/$q$-shift equation that is solvable (in this case a two term, order two $q$-shift equation). Once one object is solved, they trace back their steps to prove that $H_1(x)$ satisfies the guessed $q$-shift equation. On the other hand, at an intermediate point, by using $q$-hypergeometric transformations and summation formulas they show that $H_1(1)$ is equal to the right-hand side product of \eqref{H1_1}.

One should bear in mind that $H_1(x)$ satisfing the guessed $q$-shift equation can directly be proven using \texttt{Sigma}. Although, this would not be providing any intuition or leads towards what would be needed to reach the product side of the identity. Furthermore, in a search of a hypergeometric proof, one can use the \texttt{HYPQ} package \cite{HYPQ} to semi-automatically apply the transformation and summation formulas.

Another \texttt{qFunctions} guesser control worth mentioning here is that a user can dilate the $q$-shifts in the guessing process by the \texttt{ShiftIncrement} option. Changing the shift increment to 2 means that the system would look for a solution that only involves the functions $H(x),$ $H(xq^2)$, $H(xq^4)$, etc. One can set this parameter to any positive integer.

In the following example, we directly guess the relevant $q$-shift equation. Then we apply the first couple of substitutions that appear in the original proof of Bringmann--Jennings-Shaffer--Mahlburg. 

\begin{mma}
\In |Hse| = |GuessqShiftEquation|[(-1)^k q^{(i + 2 j + 3 k) (i + 2 j + 3 k - 1) + 3 k^2 + i + 6 j + 6 k}x^{i + 2 j + 3 k}/ \linebreak
|qPochhammer|[q, q, i]/|qPochhammer|[q^4, q^4, j]/\linebreak |qPochhammer|[q^6, q^6, k],
\{\{i, 0, 30\}, \{j, 0, 15\}, \{k, 0, 10\}\}, H[x], 3,\linebreak \{3, 12\}, |ShiftIncrement| \shortrightarrow 2]\\
\Out H[x] + (-1 - q x - q^2 x + q^3 x) H[q^2 x] - 
 q^3 x (1 - q^2 x + q^3 x + q^4 x) H[q^4 x] + 
 q^8 x^2 (-1 + q^4 x) H[q^6 x]\\
\end{mma}

Here we used an equivalent call sequence of our guesser, where the guesser is given the summand and the summation bounds instead of a list of coefficients. We also specialized the degrees of the linear system with $\{3,12\}$ in the syntax, which means that we pick the degree of $z$ to be $\leq 3$ and degree of $q$ to be $\leq 12$ in the initial setup. This is in the place of imposing a uniform bound on both variables' degrees as in the previous examples.

Now we will demonstrate the following two substitutions (one in the
$q$-shift equation level and another in the recurrence level) in the following order:
\begin{align*}
B(x) &= \sum_{n\geq 0} b_n(q) x^n  = \frac{H(x)}{(x;q^2)_\infty},\\
\intertext{and}
c(n) &= \frac{(q^2;q^2)_n}{(q;q)_{2n+1}} b(n).
\end{align*}

\begin{mma}
\In |Bsubs| = |qSESubstitute|[|Hse|, H[x], 1/|qPochhammer|[x, q^2, \infty]]]/.H \shortrightarrow B\\
\Out B[x] - \frac{((-1 - q x - q^2 x + q^3 x) B[q^2 x])}{(-1 + x)} - \frac{( q^3 x (1 - q^2 x + q^3 x + q^4 x) B[q^4 x])}{((-1 + x) (-1 + q^2 x))} - \frac{( q^8 x^2 B[q^6 x])}{((-1 + x) (-1 + q^2 x))}\\
 \In |Bse| = |Collect|[|Bsubs| // |Together| // |Numerator|, B[\_], |Factor|]\\
 \Out (-1 + x) (-1 + q^2 x) B[  x] - (-1 + q^2 x) (-1 - q x - q^2 x + q^3 x) B[q^2 x] - 
 q^3 x (1 - q^2 x + q^3 x + q^4 x) B[q^4 x] - q^8 x^2 B[q^6 x]\\
 \In brec = |qSEToqRE|[Bse, B[x], b[n]]\\
 \Out q^2 (1 + q^{1 + 2 n}) (1 + q^{2 + 2 n}) (-1 + q^{3 + 2 n}) b[n] + (1 + q^2 + q^{3 + 2 n} - q^{5 + 2 n} + q^{7 + 4 n}) b[   1 + n] + (-1 + q^{2 + n}) (1 + q^{2 + n}) b[2 + n]\\
 \In grec = |qRESubstitute|[|brec|, b[n], |qPochhammer|[q^2, q^2, n]/\linebreak
   |qPochhammer|[q, q, 2 n + 1]]/.b\shortrightarrow g\\
\Out g[n] \left(q^n-1\right) \left(q^n+1\right)+\frac{g[n-2] \left(q^{2 n}+q^2\right) \left(q^{2 n}+q^3\right) \left(\left(q^2\right)^n-1\right)
   \left(\left(q^2\right)^n-q^2\right)}{q^3 \left(q^n-1\right) \left(q^n+1\right) \left(q^n-q\right) \left(q^n+q\right) \left(q^{2 n+1}-1\right)}\linebreak -\frac{g[n-1]
   \left(q^{2 n}+q^{4 n}-q^{2 n+2}+q^3+q\right) \left(\left(q^2\right)^n-1\right)}{q \left(q^n-1\right) \left(q^n+1\right) \left(q^{2 n+1}-1\right)}\\
\In |Collect|[|grec|//|Together|//|Numerator|, g[\_], |Factor|]\\
\Out (q^2 + q^{2 n}) (q^3 + q^{2 n}) g[-2 + n] -  q^2 (q + q^3 + q^{2 n} + q^{4 n} - q^{2 + 2 n}) g[-1 + n] +  q^3 (-1 + q^n) (1 + q^n) (-1 + q^{1 + 2 n}) g[n]\\
\end{mma}

With our implemented tool set, one can easily apply and experiment within this method of guessing a relation then formally switching between $q$-shift equations and recurrences and doing substitutions to simplify these conjectural relations.

\section{The Method of Weighted Words}\label{Section_WW} 

The classical method of weighted words was first introduced by Alladi and Gordon \cite{Generalized_Schur}. This method has proven to be applicable to various problems \cite[etc.]{Alladi_Gollnitz, Alladi_Gollnitz1, Refinement, Weighted_RR, FourPar}. In the recent years, Dousse\cite[etc.]{Dousse_Unif, Dousse_Siladic, Dousse_revisit} has revisited this method and changed its main line of guessing/proving a generating function directly to forming recurrences and working on these recurrences. This gave some more flexibility to this method and increased its applicability.

The main idea of the method of weighted words is to abstractify the gap conditions between consecutive parts of partitions into distinct parts. The gaps between non-consecutive parts of partitions is not in the scope of this classical method. Here we extend the classical method from the distinct partitions to ordinary partitions. Moreover, we generalize the method to accommodate fixed number of repetitions of a part size, to allow colored partitions, etc.

We will be simultaneously explaining and demonstrating the technique using a running example. One starts with an alphabet and an ordering between letters (usually called colors), such as \[a\leq b\leq c.\]
Then one thinks of this ordering to be repeated with sub-indices to represent the level of these colors: 
\begin{equation}\label{no_weight_ordering}
a_1\leq b_1\leq c_1\leq a_2\leq b_2\leq c_2\leq a_3\leq b_3\leq c_3\leq a_4\leq b_4\leq \dots.\end{equation}
The elements in the ordering \eqref{no_weight_ordering} would affect the generating function with the weights
\begin{equation}\label{weighted_ordering}
a q, b q, c q, a q^2, b q^2, c q^2, a q^3, b q^3, cq^3, aq^4, bq^4, \dots, 
\end{equation}respectively. The sub-indices of the parts become the $q$-powers. It is easy to see that if one maps $q\mapsto q^3$, and $(a,b,c) = (1/q^2, 1/q, 1)$, the ordering \eqref{no_weight_ordering} coresponds to natural numbers and the list of weights \eqref{weighted_ordering} becomes \[ q,q^2,q^3,q^4,q^5,q^6,\dots.\] This is how one shifts from the abstract ordering of colors with levels to the natural numbers. Other weights can be used to impose other (unnatural) orders of natural numbers, to create multiple copies of a number to introduce weights, etc. 

Now one thinks of all the finite lists of non-increasing (according to the imposed order \eqref{no_weight_ordering}) colors (with indices) such as $\pi=(b_6,a_6,c_4,c_2,a_1)$. The total weight of such list is \[|(b_6,a_6,c_4,c_2,a_1)| := b q^6\, a q^6\, cq^4\, cq^2\, aq = a^2bc^2 q^{19}.\] This is analogous to the size of partitions. These lists' connection to ordinary partitions is clear by the above discussion. Thereofer, We will start calling these lists \textit{partitions}.

One can now impose any arbitrary set of rules on these partitions, such as \begin{quotation}if $a_n$ is a part then $c_{n-1}$ is not a part of the partition, or\newline $b_n$ cannot appear more than once as a part of a partition, etc.\end{quotation} The collection of these rules can be represented as a square matrix that encodes these gap (of index) rules for all the possible combinations of letters in its entries. One example gap matrix is the following \begin{equation}\label{Schur_matrix}
\mathbf{M} = \begin{blockarray}{cccc}
 & a & b & c \\
\begin{block}{c(ccc)}
  a & 1 & 2 & 2 \\
  b & 1 & 1 & 2 \\
  c & 1 & 1 & 2 \\
\end{block}
\end{blockarray}
\end{equation}

The interpretation of $\mathbf{M}_{i,j}$ is the least permissible gap condition for going from $k$-th level of the letter in the $i$-th row to the letter in the $j$-th column. In \eqref{Schur_matrix}, the $\mathbf{M}_{1,3}=2$ encodes that in these gap conditions it is permissible to have $a_k$ and $c_{k-2}$ (but not with $c_{k-1}$) in the partitions. The $\mathbf{M}_{2,2}=1$ encodes that $b_k$ can appear together with $b_{k-1}$ (but not with another $b_k$).

Let $g_{a_k}(a,b,c)$, $g_{b_k}(a,b,c)$ and $g_{c_k}(a,b,c)$ be generating functions for all the weights of partitions that satisfy the gap conditions mandated by $\mathbf{M}$ \eqref{Schur_matrix}, where parts are $\leq a_k$, $\leq b_k$ and $\leq c_k$, respectively. Observe that in the limit $k\rightarrow\infty$ these generating functions are the same :\begin{equation}\label{WW_k_inf}
g_\infty (a,b,c) = \lim_{k\rightarrow \infty}g_{a_k}(a,b,c) = \lim_{k\rightarrow \infty} g_{b_k}(a,b,c) = \lim_{k\rightarrow \infty} g_{c_k}(a,b,c).
\end{equation}

A couple of  initial values for these generating functions are given in Table~\ref{WW_initial_conds}.

\begin{table}[h]\caption{Initial values for $g_{a_k}(a,b,c)$, $g_{b_k}(a,b,c)$, and $g_{c_k}(a,b,c)$.}\label{WW_initial_conds}
$\begin{array}{l}
g_{a_1}(a,b,c) = 1 + a q,\\
g_{a_2}(a,b,c) = 1+ aq + bq + cq + aq^2(1+aq),\\[-1.8ex]\\
g_{b_1}(a,b,c)=1+ aq + bq,  \\
g_{b_2}(a,b,c) =  1+ aq + bq + cq + aq^2(1+aq) + b q^2 ( 1+ aq + bq),\\[-1.8ex]\\
g_{c_1}(a,b,c)=1+ aq + bq+ cq,\\
g_{c_2}(a,b,c) = 1+ aq + bq + cq + aq^2(1+aq) + b q^2 ( 1+ aq + bq) + c q^2( 1+ aq + bq).\\
\end{array}$
\end{table}

One can use the gap matrix $\mathbf{M}$ to form recurrences for $g_{a_k}(a,b,c)$, $g_{b_k}(a,b,c)$ and $g_{c_k}(a,b,c)$. These recurrences are formed by instance considerations. For the recurrence $g_{a_k}(a,b,c)$, assuming $k$ is large enough (larger than the maximum of the entries of the gap matrix $\mathbf{M}$), one considers the following:
\begin{quotation}
If $a_k$ is not a part, the next largest possible part size that may appear is $c_{k-1}$. If $a_k$ is a part (and changes the overall weight by $a q^k$) the largest permissible part is $a_{k-1}$. Moreover, anything smaller than $a_{k-1}$ is also permissible. 
\end{quotation}
This translates to recurrences as \begin{equation}\label{g_a_rec}g_{a_k}(a,b,c) = g_{c_{k-1}}(a,b,c) + a q^k g_{a_{k-1}}(a,b,c).\end{equation} The other recurrences can be found in a similar manner. 

The recurrence system for a gap matrix can automatically be found (using the same type of considerations while forming the recurrences) by the \texttt{qFunctions} package:
\begin{mma}
\In |recs| = |GenerateRecurrencesFromMatrix|[\{\{1, 2, 2\}, \{1, 1, 2\}, \{1, 1, 2\}\}, \{a,  b, c\}, g[k]]\\
\Out \{-a q^k g[a][-1 + k] + g[a][k] - g[c][-1 + k], -g[a][k] -   b q^k g[b][-1 + k] + g[b][k],\newline -c q^k g[b][-1 + k] - g[b][k] +  g[c][k]\}\\
\end{mma}

This is done systematically and it is equivalent to finding these recurrences by hand. Here the syntax $g[a][k]$ represents the function $g_{a_k}(a,b,c)$, and the rest is defined similarly. The first recurrence in {\texttt{Out [19]}} is the same recurrence as \eqref{g_a_rec}, as in the \texttt{qFunctions} outputs we set any and all recurrences equal to 0. 

Dousse's spin on this method comes into play at this point. She tries to find a functional relation between two different families of generating functions, using substitutions and index shifts. Finding functional equations this way requires ingenuity and good intuition. We would like to simplify this problem slightly by uncoupling the recurrences. This way maybe one can see which generating functions satisfy similar recurrence relations and what substitutions might be useful.

There are numerous implemented machineries to uncouple a coupled system of recurrences. The \texttt{qFunctions} package uses the \texttt{HolonomicFunctions} package of Koutschan \cite{HolonomicFunctions}. One needs to import this package before asking for uncoupling of recurrences.

\begin{mma}
\In \mathbf{< \hspace{-1.5mm}< } |RISC`HolonomicFunctions`|\\
\Print \vspace{-.2cm}\\
\Print \fbox{HolonomicFunctions Package by Christoph Koutschan - $\copyright$ RISC Linz - V. 1.7.3 (21-Mar-2017)}\\
\In |uncoupledsRecs|=|UncoupleRecurrences|[|recs|,\{g[a],g[b],g[c]\},k]\\
\Out \{(-c q^{2 + k} + a b q^{4 + 2 k}) g[a][    1 + k] + (-1 - b q^{2 + k} - a q^{3 + k}) g[a][2 + k] +   g[a][3 + k],\newline (c q^{1 + k} - a b q^{3 + 2 k}) g[b][    k] + (1 + a q^{2 + k} + b q^{2 + k}) g[b][1 + k] -   g[b][2 + k],\newline (-c q^{3 + k} + a b q^{5 + 2 k}) g[c][
    1 + k] + (-1 - a q^{3 + k} - b q^{3 + k}) g[c][2 + k] +   g[c][3 + k]\}\\
\end{mma}

In this example, it is clear that these recurrences are looking quite alike in their uncoupled form, maybe with a sign difference. If we focus on the last two recurrences, the polynomial coefficients of these recurrences become the same if one maps $c\mapsto a b q$ in both recurrences and then $(a,b)\mapsto(aq,bq)$ in the recurrence of the $g_{b_k}(a,b,c)$. In other words, $g_{b_k}(aq,bq,ab q^3)$ and $g_{c_{k+1}}( a,b,abq)$ satisfy the recurrence \begin{equation}\label{SchurRec}(a b q^{4 + k} - a b q^{5 + 2 k})a(k) + (1 + a q^{3 + k} + b q^{3 + k})a(k+1) -  a(k+2)=0.\end{equation} Note that the used substitution is not the only one or the simplest one that would make the coefficients of these two recurrences of $g_{b_k}(a,b,c)$ and $g_{c_k}(a,b,c)$ the same.

One can now examine some initial values of $g_{b_k}(aq,bq,a b q^3)$ and $g_{b_k}( a,b,ab q)$ in search of any relation between these two generating functions. We list the $k=1,2,3$ values for both functions.

\begin{mma}
\In \{|gbRec|,|gcRec|\} = \{|uncoupledsRecs|[[2]],|uncoupledsRecs|[[3]]\}/.\{g[b]\shortrightarrow|gb|, g[c]\shortrightarrow|gc|\};\\
\In |qREToList|[|gbRec|, |gb|[k], \{1, \{1 + a q + b q, 1 + a q + b q +  c q + a q^2 (1 + a q) + b q^2 (1 + a q + b q)\} /. c \shortrightarrow a b q /. a \shortrightarrow a q /.  b \shortrightarrow b q\}, 3 ] // |Factor| // |TableForm|\\
\Out 1 + a q^2 + b q^2, \newline
 1 + a q^2 + b q^2 + a q^3 + b q^3 + a b q^4 + a^2 q^5 + a b q^5 + 
  b^2 q^5,\newline 1 + a q^2 + b q^2 + a q^3 + b q^3 + a q^4 + b q^4 + 
  a b q^4 + c q^4 + a^2 q^5 + a b q^5 + b^2 q^5 + a^2 q^6 + 
  2 a b q^6 + b^2 q^6 + a c q^6 + b c q^6 + a^2 q^7 + a b q^7 + 
  b^2 q^7 + a^2 b q^8 + a b^2 q^8 + a^3 q^9 + a^2 b q^9 + a b^2 q^9 + 
  b^3 q^9\\
  
\In |qREToList|[|gcRec|, |gc|[k], \{1, \{1 + a q + b q + c q, 1 + a q + b q +  c q + a q^2 (1 + a q) + 
 b q^2 (1 + a q + b q) + c q^2 (1 + a q + b q)\} /. c \shortrightarrow a b q \}, 3] // |Factor| // |TableForm|\\
\Out(1 + a q) (1 + b q),\newline (1 + a q) (1 + b q) (1 + a q^2 + b q^2),\newline (1 + 
    a q) (1 + b q) (1 + a q^2 + b q^2 + a q^3 + b q^3 + c q^3 + 
    a^2 q^5 + a b q^5 + b^2 q^5)\\
\end{mma}

The initial conditions used are the ones in Table~\ref{WW_initial_conds} and we do the mentioned substitutions on the spot. This clearly shows that \begin{equation}\label{WW_func_rel}  g_{c_{k+1}} ( a,b,abq) = (1+a q)(1+ b q) g_{b_{k}}(aq,bq,ab q^3),\end{equation} for $k=1$ and $2$. Furthermore, this is enough to prove \eqref{WW_func_rel} for all $k\geq 1$ since the generating functions satisfy the same linear recurrence \eqref{SchurRec} and the necessary initial conditions of both sides match.

Once such a relation is established, one tends $k\rightarrow\infty$ to get \[g_\infty( a,b,abq) = (1+a q)(1+ b q) g_\infty( aq,bq,ab q^3) \] and then iterates this relation \begin{align*}
g_\infty( a,b,abq) &= (1+a q)(1+ b q) g_\infty( aq,bq,ab q^3),\\
 &=(1+a q)(1+ b q)(1+aq^2)(1+b q^2) g_\infty( aq^2,bq^2,ab q^5),\\
 &\hspace{.2cm}\vdots \\
 &= (-aq;q)_\infty (-bq;q)_\infty \lim_{n\rightarrow\infty} g_\infty( aq^{n},bq^{n},ab q^{2n+1}).
 \intertext{This is the only location that we explicitly require $|q|<1$. With that we have} 
 &=(-aq;q)_\infty (-bq;q)_\infty\, g_\infty( 0,0,0). 
 \end{align*}
It is clear that $g_\infty( 0,0,0)=1$ and this yields the final result:
\begin{equation}\label{WW_Schur_prod} g_\infty( a,b,abq) = (-aq;q)_\infty (-bq;q)_\infty.\end{equation}

The identity \eqref{WW_Schur_prod} is the abstract/general version of many gap-congruence identities, that is originally due to Alladi--Gordon \cite{Generalized_Schur}. The proof technique used is a variation of Dousse's proofs. In addition to the automated creation and uncoupling of the recurrences, the novelty of the proof is that the choice of the gap matrix and the substitutions are different from previous proofs. 

To summarize, we start with some gap conditions imposed on partitions. We show, using the recurrences and substitutions, that, in the limit, the infinite (unbounded) generating function that counts the partitions with some specific weights (in this example $(a,b,c) = (a,b,abq)$), has a product representation. These product representations involving $q$-Pochhammer symbols are in direct relation with congruence conditions. Therefore, we start with a gap condition and show that the partitions counted by these gap conditions are the same as the number of partitions counted by the congruence conditions that the products impose. 

\subsection{Extensions of the method of weighted words}

As mentioned in the beginning of the section, the original method of weighted words is defined for the partitions into distinct parts. For a gap matrix $\mathbf{M}$ this translates to the upper triangular portion of the matrix having positive values. In other words  \[\mathbf{M}_{i,j}\geq 1\] for all $i\geq j$. In \texttt{qFunction}, we allow 0's on the main diagonal to indicate that a part size can repeat. This extends the original  method to ordinary partitions. 

We also go beyond the ordinary partitions by allowing 0's in the strictly upper triangular portion of the matrix. This convention breaks the ordering between the letters. We remind the reader that there are two orders in play, one order is the order of the suc-indices which always stays intact, and the other is the introduced order of type $a\leq b\leq c$ between the letters. By letting 0's to appear on the strictly upper triangular of the gap matrix the order between the letters can be relaxed to no order relation between the terms. An example of this is as follows:

\begin{mma}
\In |GenerateRecurrencesFromMatrix|[\{\{0, 0\}, \{0, 0\}\}, \{a, b\}, g[k]]\\
\Out \{g[a][k] - g[b][-1 + k] - a q^k g[b][k], -g[a][k] - \frac{b q^k g[a][k]}{  1 - b q^k} + g[b][k]\}\\
\In |UncoupleRecurrences|[\%, \{g[a], g[b]\}, k]\\
\Out \{(1 - b q^{2 + k}) g[a][1 + k] + (-1 + b q^{1 + k} + a q^{2 + k} + b q^{2 + k} - 
     a b q^{3 + 2 k} - b^2 q^{3 + 2 k}) g[a][2 + k],\newline -g[b][ k] + (1 - a q^{1 + k} - b q^{1 + k}) g[b][1 + k]\}\\
\end{mma}

Hence, with the initial condition $g_{b_1} = 1+a q+ b q$, we have \begin{align}\nonumber g_{b_k}(a,b) &= \prod_{i\geq 1} \frac{1}{(1-aq^i -bq^i)}\\ \label{WW_weighted_list}&= 1+ (a+b)q+ (a + a^2 + b + 2 a b + b^2)q^2\\\nonumber&\hspace{1cm}+(a + a^2 + a^3 + b + 2 a b + 3 a^2 b + b^2 + 3 a b^2 + b^3) q^3+\dots.\end{align} On the other hand, interpretation of the gap matrix-wise, this is the generating function of weights of partitions where any ordering of $a$ and $b$ can appear. As an example all the partitions that come with $q^3$ weight are listed as follows:
\[\begin{array}{c}(a_3),(a_2,a_1), (a_1,a_1,a_1),(b_3),(a_2,b_1),(b_2,a_1),(a_1,a_1,b_1),(a_1,b_1,a_1),\\[-1.5ex]\\ (b_1,a_1,a_1),(b_2,b_1),(a_1,b_1,b_1),(b_1,b_1,a_1),(b_1,a_1,b_1),(b_1,b_1,b_1),\end{array}\] respective to the coefficients in the series \eqref{WW_weighted_list}.

Lastly, we extend the method of weighted words to allow the condition that a part can appear up to $n$ times. This is done with the convention that one enters $-n$ to in the diagonal entry of the gap matrix. That being said $-1$ as a diagonal entry is redundant and it is the same as having a 1 in that diagonal entry. For example, given the alphabet order $a\leq b$, we encode the partitions, where $a_k$'s appear at most 3 times and $b_k$'s appear at most $4$ times with the matrix \[\begin{blockarray}{ccc}
 & a & b  \\
\begin{block}{c(rr)}
  a & -3 & 1  \\
  b & 0 & -4  \\
\end{block}
\end{blockarray}.\]
The related weighted words recurrences are found automatically in the same manner:

\begin{mma}
\In |GenerateRecurrencesFromMatrix|[\{\{-3, 1\}, \{0, -4\}\}, \{a, b\}, g[k]]\\
\Out \{g[a][k] - g[b][-1 + k] - \frac{a q^k (1 - a^3 q^{3 k}) g[b][-1 + k]}{
  1 - a q^k}, -g[a][k] - \frac{b q^k (1 - b^4 q^{4 k}) g[a][k]}{ 1 - b q^k} + g[b][k]\}\\
\end{mma}

Notice that we assume that $b_k$ can come with $b_{k-1}$ in this situation. If one would require a set-up that would both allow $b_k$ to repeat up to some certain number of times and not appear together with $b_{k-i}$ for some positive integer $i$, then one can always increase the alphabet size, introduce new letters, and achieve this. As an example if one would like to put the extra condition that $b_k$ and $b_{k-1}$ do not come together in the last example, we can dilate our alphabet two-fold and have $a\leq b\leq a' \leq b'$ and put the condition that $b'_k$ does not appear with $b_k$ and $b_k$ does not appear with $b'_k$ in the gap matrix.

The authors hope that with these extensions on the original method some of the conjectures of Kanade--Russell~\cite{Kanade_Russell, IdentityFinder, Mathew_Thesis} or the ones of Nandi~\cite{Nandi} would go under the weighted words umbrella, and become theorems. 

\subsection{Warnings and termination} Not every matrix can be used as a gap matrix, or would yield a finite order recurrence as a gap matrix. These type of matrices are identified and automatically ignored by \texttt{qFunctions}. As an example 

\begin{mma}
\In |GenerateRecurrencesFromMatrix|[\{\{1,1,1\}, \{2, 1,1\},\{0,0,1\}\}, \{a, b,c\}, g[k]]\\
\Print Given difference conditions matrix either does not yield a finite order recurrence, or the difference conditions are inconsistent.\\
\Out \{\}\\
\end{mma}

In this example the gap conditions are not consistent. The matrix \[\mathbf{M} = \begin{blockarray}{cccc}
 & a & b & c \\
\begin{block}{c(ccc)}
  a & 1 & 1 &1 \\
  b & 2 & 1 & 1 \\
  c & 0&0&1\\
\end{block}
\end{blockarray}\] suggests (by looking at $\mathbf{M}_{2,1} = 2$) that if $b_{k}$ is a part of a partition then $a_{k-1}$ is not. On the other hand $b_{k}$ can be together with $c_{k-1}$ (gap condition at $\mathbf{M}_{2,3} = 1$) and $c_{k-1}$ can come together with $a_{k-1}$  (gap condition at $\mathbf{M}_{2,3} = 0$). Together, this creates a discrepancy in the gap conditions and violates the gap condition $\mathbf{M}_{2,1}$. Thus, we choose to discard such matrices, and terminate the process of forming recurrences. This way we can guarantee that regardless of the user input the process terminates.

\section{A scheme on cylindric partitions}\label{Section_Cylindric}

Cylindric partitions were introduced by Gessel--Krattenthaler \cite{Cylindric}: 
\begin{definition}For $k\in \mathbb{N}$, and a composition $C=(c_1,c_2,\dots,c_k)$ with $c_i\geq 0$, a cylindric partition of profile $C$ is a vector partition $(\pi^{1},\pi^2,\dots, \pi^k)$, which satisfies the properties \[\pi^i_{j}\geq \pi^{i+1}_{j+c_{i+1}}\text{ and }\pi^k_{j}\geq \pi^{1}_{j+c_{1}}\] for all $i$ and $j$.
\end{definition}

It is noted that $(2,1,1)$, $(1,2,1)$, and $(1,1,2)$ are all the same as profiles for cylindric partitions, and in general these types of cyclic shifts in the profile does not change the counted partitions. Given a profile $C=(c_1,c_2,\dots,c_k)$, we call $C'=(c'_1,c'_2,\dots,c'_k)$ an \textit{adjacent profile} if it has the same number of elements and the sizes are the same $|C| = c_1+c_2+\dots+c_k = c'_1+c'_2+\dots+c'_k =|C'|$.

Let $\P_C$ be the set of all cylindric partitions for a given profile $C$. We define the two variable generating function for these cylindric partitions as \[F_C(z) = F_C(z,q) := \sum_{\pi\in\P_C} q^{|\pi|} z^{l(\pi)},\] where $|\pi|$ is defined as the sum of all the partitions in the vector partition $\pi$ and $l(\pi)$ is the largest part size among the partitions of $\pi$. 

Then Borodin's beautiful theorem states the following \cite{Borodin}.

\begin{theorem}[Borodin, 2007]\label{Thm_Borodin} For a given profile $C=(c_1,c_2, \dots,c_k)$ let $t:= k + |C|$, then
\begin{equation}\label{Borodin_prod}F_C(1) = \frac{1}{(q^t;q^t)_\infty} \prod_{i=1}^k \prod_{j=i}^k\prod_{m=1}^{c_i} \frac{1}{(q^{m+s(i+1,j)+j-i};q^t)_\infty}\prod_{i=2}^k \prod_{j=2}^{i}\prod_{m=1}^{c_i} \frac{1}{(q^{t-m+s(j,i-1)+j-i};q^t)_\infty},  \end{equation}
where $s(i,j) = c_i+c_{i+1} +\dots + c_{j-1}+c_j$.
\end{theorem}

Corteel and Welsh \cite{Corteel} wrote explicit functional equations for $F_C(z)$ and \begin{equation}\label{G_g_def}G_C(z) := G_C(z,q) =\sum_{n\geq 0} g_C(n,q) z^n = (zq;q)_\infty F_C(z).\end{equation} We will be using the abbreviation $g_C(n)$ for $g_C(n,q)$. In the functional equations $F_C(z)$ gets related with functions of adjacent profiles. This is a coupled system of $q$-shift equations. This relation is formulated combinatorially over the location of the possible maximum part size in the cylindric partitions. 

It is clear that finding the form of the $F_C(y)$ together with Theorem~\ref{Thm_Borodin} would yield sum-product identities. This is a difficult task. The only profiles that have been succesfully studied through this scheme (to the best of our knowledge) are the profiles $(2k+1-i,i)$ by Foda-Welsh \cite{Foda}, for $2k+1\geq i\geq 0$, and $(4,0,0)$ and its adjacent profiles by Corteel-Welsh \cite{Corteel}. 

The \texttt{qFunctions} package includes the creation of the system of equations for a family of adjacent profiles. Here we demonstrate this on the profiles with $2$ parts, and present some sum-product identities using a result by Hirschhorn \cite{Hirschhorn}.

\begin{mma}
\In |sys| = |CylindricalFunctionalEquationSystem|[G,\{2, 4\}, z, q]\\
\Out \{G[\{2, 2\}][z] + (1 - q z) G[\{2, 2\}][q^2 z] -  2 G[\{3, 1\}][q z],\newline
 -G[\{3, 1\}][q z] + G[\{4, 0\}][z], \newline
-G[\{2, 2\}][q z] +  G[\{3, 1\}][z] + (1 - q z) G[\{3, 1\}][q^2 z] - G[\{4, 0\}][q z]\}\\
\In |coeffSys| = |FunctionalEquationToRecurrences|[|sys|, G[z, q], g[n]]\\
\Out \{-q^{-1 + 2 n} g[\{2, 2\}][-1 + n] + (1 + q^{2 n}) g[\{2, 2\}][n] -  2 q^n g[\{3, 1\}][n],\newline -q^n g[\{3, 1\}][n] + 
 g[\{4, 0\}][n], \newline -q^n g[\{2, 2\}][n] -  q^{-1 + 2 n} g[\{3, 1\}][-1 + n] + (1 + q^{2 n}) g[\{3, 1\}][n] - 
 q^n g[\{4, 0\}][n]\}\\
\end{mma}

Here the syntax $G[\{2, 2\}][z]$ and $g[\{2, 2\}][n]$ are referring to $G_{(2, 2)}(z,q)$ and $g_{(2, 2)}(n)$, respectively. The rest of the functions are defined in the same fashion.

This system can be uncoupled as in Section~\ref{Section_WW}, and then be studied and simplified using the techniques discussed in Section~\ref{Section_Def}. Here we use the auxillary function \texttt{CoefficientListMaker} which lists all the possible names for the coefficients of $G_C(z)$ functions in the given profile family. One can also use \texttt{FunctionListMaker} in the same fashion to list the names of the series names in a profile family.

\begin{mma}
\In  |coefflist| = |CoefficientListMaker|[g, \{2, 4\}]\\
\Out \{g[\{2, 2\}], g[\{3, 1\}], g[\{4, 0\}]\}\\
\In |UncoupSys| =  |UncoupleRecurrences|[|coeffSys|,|coefflist|, n]\\
\Out \{q^{5 + 4 n} g[\{2, 2\}][ n] + (-q^{3 + 2 n} - q^{4 + 2 n} - q^{6 + 4 n}) g[\{2, 2\}][    1 + n] + (1 - q^{4 + 2 n}) g[\{2, 2\}][2 + n],\newline -q^{5 + 4 n} g[\{3, 1\}][ n] + (q^{3 + 2 n} + q^{4 + 2 n} + q^{7 + 4 n}) g[\{3, 1\}][  1 + n] + (-1 + q^{4 + 2 n}) g[\{3, 1\}][2 + n], \newline -q^{11 + 4 n} g[\{4, 0\}][1 + n] + (q^{6 + 2 n} + q^{7 + 2 n} + q^{12 + 4 n}) g[\{4, 0\}][ 2 + n] + (-1 + q^{6 + 2 })) g[\{4,0\}][3 + n]\}\\
\end{mma}

Now we take the first recurrence and do the substitution \begin{equation}\label{g_h_substitution}h_{(2,2)} (n) =g_{(2,2)}\frac{(q^2;q^2)_n}{q^{n^2}}.\end{equation} One can see that this substitution naturally arises by listing some values of the $g_{(2,2)}(n)$ with the initial conditions $g_{(2,2)}(0)=1$ and $g_{(2,2)}(-1)=0$.

\begin{mma}
\In |rec1| = |UncoupSys|[[1]]/.g[\{2, 2\}]\shortrightarrow g22\\
\Out q^{5 + 4 n} g22[ n] + (-q^{3 + 2 n} - q^{4 + 2 n} - q^{6 + 4 n}) g22[    1 + n] + (1 - q^{4 + 2 n}) g22[2 + n]\\
\In |rec2| = |qRESubstitute|[rec, g22[n], |qPochhammer|[q^2, q^2, n]/q^{n^2}]/.g22\shortrightarrow h22\newline//   |PowerExpand| // |ExpandAll|;\\
\In |Collect|[|rec2| // |Together| // |Numerator|, h22[\_], |Factor|]\\
\Out q^{5 + 4 n} (-1 + q^{1 + n}) (1 + q^{1 + n}) h22[n] +  q^{4 + 4 n} (1 + q + q^{3 + 2 n}) h22[1 + n] - q^{4 + 4 n} h22[2 + n]\\
\end{mma}

The recurrence for the $h_{(2,2)}(n)$ in {\small \texttt{Out [35]}} is equivalent to the recurrence \begin{equation}
\label{rec_2_2} a_{n+2} = (1+ q + q^{2n+3}) a_{n+1} + q(-1+q^{2n+2}) a_{n}.
\end{equation} The sequence $h_{(2,2)}(n)$ satisfies (\ref{rec_2_2}) with the initial conditions $h_{(2,2)}(-1)=0$ and $h_{(2,2)}(0)=1$. This recurrence with these initial conditions was solved in a higher generality by Hirschhorn \cite{Hirschhorn}.

\begin{theorem}[Hirschhorn, 1974]For positive integer $n$,
\begin{equation}\label{Hirschhorn_sum}\sum_{i,j,k\geq0} a^i b^{n-i-j-k}c^jd^{k-j}q^{\binom{j+1}{2}+\binom{k+1}{2}}{k+i\brack i}_q{n-i-j\brack k}_q{k\brack j}_q\end{equation}
is the solution of the recurrence relation
\[a_{n+2}= (a+b+dq^{n+2})a_{n+1}+(-ab+cq^{n+2})a_{n}\] with the initial conditions $a_{-1}=0 $ and $a_0=1$.
\end{theorem}

Whence, \begin{equation}\label{h_Hirschhorn}h_{(2,2)}(n) = \sum_{i,j,k\geq0} q^{j^2+k^2+n-i-k}{k+i\brack i}_{q^2}{n-i-j\brack k}_{q^2}{k\brack j}_{q^2}.\end{equation} This is what one gets by $q\mapsto q^2$ and $(a,b,c,d) = (1,q,q^{-1},q^{-1})$ in \eqref{Hirschhorn_sum}. Tracing back the substitutions \eqref{g_h_substitution} and \eqref{G_g_def}, we get that \begin{equation}
\label{G_2_2} G_{(2,2)}(z) = \sum_{i,j,k,n\geq 0 } \frac{q^{n^2+j^2+k^2+n-i-k}}{(q^2;q^2)_n}{k+i\brack i}_{q^2}{n-i-j\brack k}_{q^2}{k\brack j}_{q^2} z^n
\end{equation}

Finally, going back to $F_{(2,2)}(z)$, setting $z=1$ to use Theorem~\ref{Thm_Borodin}, and simplifying the identity yields the following sum-product formula.

\begin{theorem}\label{Thm_profile22} We have
\begin{equation}
\sum_{i,j,k,n\geq 0 } \frac{q^{n^2+j^2+k^2+n-i-k}}{(q^2;q^2)_n}{k+i\brack i}_{q^2}{n-i-j\brack k}_{q^2}{k\brack j}_{q^2} = \frac{(q^3;q^6)_\infty}{(q,q^2,q^4,q^5;q^6)_\infty}.
\end{equation}
\end{theorem} 

To ease the evaluation of the product \eqref{Borodin_prod} we have included the \texttt{BorodinProductPrinter} function to \texttt{qFunctions}. One can directly ask for the outcome of Theorem~\ref{Thm_Borodin} with any given cylindric partition profile and a variable~$q$.

\begin{mma}
\In |BorodinProductPrinter|[\{2, 2\}, q]\\
\Out \frac{1}{(q,q,q^2,q^2,q^4,q^4,q^5,q^5,q^6;q^6)_\infty}\\
\end{mma}

Similar to Theorem~\ref{Thm_profile22}, we can prove the following two results for $F_{(3,1)}(1)$ and $F_{(4,0)}(1)$, respectively. These results are both related to \eqref{Hirschhorn_sum} with $q\mapsto q^2$ and $(a,b,c,d) = (1,q,q^{-1},1)$.

\begin{theorem}\label{Thm_prof_40_31} We have,
\begin{align*}
\sum_{i,j,k,n\geq 0 } \frac{q^{n^2+j^2+k^2+n-i-j}}{(q^2;q^2)_n}{k+i\brack i}_{q^2}{n-i-j\brack k}_{q^2}{k\brack j}_{q^2} &= \frac{1}{(q,q^3,q^5;q^6)_\infty},\\
\sum_{i,j,k,n\geq 0 } \frac{q^{n^2+j^2+k^2+2n-i-j}}{(q^2;q^2)_n}{k+i\brack i}_{q^2}{n-i-j\brack k}_{q^2}{k\brack j}_{q^2} &= \frac{1}{(q^2,q^3,q^4;q^6)_\infty}.
\end{align*}
\end{theorem}

Notice that the products in Theorems~\ref{Thm_profile22} and \ref{Thm_prof_40_31} are the ones that appear in Bressoud's products. It is highly likely that the sum sides can be simplified to the sum sides of Bressoud's theorem. This reduction/summation exercise is left to interested readers.

\section{\texorpdfstring{$q$}{q}-Fitting}\label{Section_q_fit}

In the previous sections, it became a running theme that we end up with a recurrence defining a function, which we don't know the formula of. It is clear that we can make a table of values for these functions using necessary initial conditions and the defining recurrence. If only we can guess an explicit formula of the function, it is highly likely that we can prove that the suggested function satisfies the defining recurrence using symbolic computation tools, such as \texttt{Sigma} \cite{Sigma}, \texttt{qMultiSum} \cite{qMultiSum}, etc. 

Therefore, it is relevant to have a fitting functionality in the \texttt{qFunctions} package, where one can try to write a given data set as a sum of simple expressions, such as $q$-binomials, $q$-trinomials, etc. A similar implementation already exists in Maple as a part of the \texttt{RRtools} package by Sills \cite{Sills_RRtools,Sills_FinRR}. To avoid any confusion, once again, we note that the definitions of $q$-trinomial's used in \texttt{RRtools} are due to Andrews and Baxter \cite{Andrews_Baxter}, whereas we are following Warnaar \cite{Warnaar} for the definitions of the $T_n$ and $t_n$. This is a small dilation difference of $q\mapsto q^2$ as mentioned in Section~\ref{Section_Def}.

Given a list of polynomials, one can try to understand if these polynomials, or an arithmetic subset of these polynomials, can be written as a sum in $q$-binomials, $q$-trinomials, etc. This can be done with an exhaustive search and it is implemented in the \texttt{qFunctions} package as the \texttt{FitqRepresentation} function. One can also use the more specialized functions such as \texttt{FitqVnTrinomialRepresenation}, but this can also be done by changing the \texttt{TryToFit} option of the \texttt{FitqRepresentation} to the desired function name (as an example, to \texttt{"qVnTrinomial"}). \texttt{TryToFit} option is set to \texttt{"All"} by default. Moreover, \texttt{FitqRepresentation} stops its search once a representation is found, but one can use the option \texttt{ReturnAll$\shortrightarrow$True} to find different fitting polynomials for the same list. 

Let $a_n$ be a sequence that satisfies the recurrence \eqref{rec_2_2} with the initial conditions $a_{-1}=0$ and $a_0=1$. We demonstrate fitting a list of values (for the non-negative indices of $a_n$) generated by the sequence as follows:

\begin{mma}
\In rec =   a[2 + n]-(1 + q + q^{3 + 2 n}) a[1 + n]-q(-1 + q^{2 +2n}) a[n]; \\
\In |aList| = |qREToList|[|rec|, a[n], \{-1, \{0, 1\}\}, 11] // |Simplify|;\\
\In |aListTrim| = |Take|[ |aList|, \{2,11\}] \\
\Out \{1,1 + 2 q, 1 + 2 q + 2 q^2 + 2 q^3 + 2 q^4,1 + 2 q + 2 q^2 + 4 q^3 + 4 q^4 + 4 q^5 + 4 q^6 + 2 q^7 + 2 q^8 + 
 2 q^9,\newline \dots,\newline \dots  + 30 q^{69} + 24 q^{70} + 20 q^{71} + 16 q^{72} + 12 q^{73} + 
 10 q^{74} + 8 q^{75} + 6 q^{76} + 4 q^{77} + 4 q^{78} + 2 q^{79} + 2 q^{80} +  2 q^{81} \}\\
 \In |FitqRepresentation|[|aListTrim|, n, q]\\
 \Out |qBinomial|[1 + 2 n, 1 + n, q] + (q - q^2) |qBinomial|[1 + 2 n, 2 + n, q] \newline - 
 q^5 |qBinomial|[1 + 2 n, 3 + n, q] - q^7 |qBinomial|[1 + 2 n, 4 + n, q]\newline + (-q^{12} + q^{15}) |qBinomial|[ 1 + 2 n, 5 + n, q] + q^{22} |qBinomial|[1 + 2 n, 6 + n, q] \newline +  q^{26} |qBinomial|[1 + 2 n, 7 + n, q]+ (q^{35} - q^{40}) |qBinomial|[1 + 2 n, 8 + n, q] \newline - q^{51} |qBinomial|[1 + 2 n, 9 + n, q] - q^{57} |qBinomial|[1 + 2 n, 10 + n, q]\\
\end{mma}

This fitted \texttt{qBinomial} polynomial and its coefficients suggests the following result:

\begin{theorem} Let $h_{(2,2)}(n)$ be defined as \eqref{g_h_substitution}, then
\begin{equation}\label{h_fitt} h_{(2,2)}(n) = \sum_{k=-\infty}^\infty {2n+1 \brack n+1+3k}(-1)^k q^{k(6k+1)} + \sum_{k=-\infty}^\infty {2n+1 \brack n+2+3k}(-1)^k q^{(2k+1)(3k+1)}.
\end{equation}
\end{theorem}

Comparing \eqref{h_fitt} to \eqref{h_Hirschhorn} shows the extend of this simplification. The proof of this result is an easy application of existing symbolic computation implementations. One finds the recurrences satisfied by the sums in \eqref{h_fitt} by one of the packages \texttt{HolonomicFunctions} \cite{HolonomicFunctions}, \texttt{Sigma} \cite{Sigma}, \texttt{qMultiSum} \cite{qMultiSum}, \texttt{qZeil} \cite{qZeil}, etc. Then one finds the greatest common divisor of these recurrences and the defining recurrence \eqref{rec_2_2} (the g.c.d. turns out to be \eqref{rec_2_2} itself). Finally, checking the initial conditions finishes the proof of \eqref{h_fitt}.

We demonstrate fitting the terms $a_{3n}$ in the sequence with either \texttt{qTrinomial}s or \texttt{qTnTrinomial}s:
 
\begin{mma}
 \In |FitqRepresentation|[|aListTrim|, n, q, \{3,0\}, |TryToFit|\shortrightarrow \{"|qTrinomial|","|qTnTrinomial|"\}]\\
 \Out |qTrinomial|[1 + 2 n, 1 + n, 1 + n,  q] + (q + q^2 + 4 q^3 + 4 q^4 + 4 q^5 + 4 q^6 + 2 q^7 + 2 q^8 + 
    2 q^9) |qTrinomial|[1 + 2 n, 1 + n, 2 + n, q] + (-2 q^2 - 4 q^3 - \dots  + 2 q^{35} + 2 q^{36}) |qTrinomial|[1 + 2 n, 1 + n, 3 + n, q]\\
\end{mma} 

Where the $\{3,0\}$ in the syntax is suggesting that we are only looking at the 0 modulo 3 indexed terms in the given list. Moreover, the functions tried in fitting the said polynomials are the \texttt{qTrinomial}s and the \texttt{qTnTrinomial}s, defined in \eqref{Round_Tri_Def} and \eqref{T_n_Def}, respectively. Among these options the function successfully finds polynomials to fit the data using  \texttt{qTrinomial}s, and ends the calculations there. Therefore, that solution is represented as output of the function. 

\section{Final Words and Prospects}\label{Section_Outlook}

We are hoping that this package will be used by many, and that it will be put to good use. This first version of the \texttt{qFunctions} package also includes some simple \texttt{qseries} package inspired tools such as \texttt{qProdMake}. The interested users are encouraged to look at the help documentation of \texttt{qFunctions} of those non-mentioned functions at their own time. 

A long run goal is the seamless integration of this package with \texttt{Sigma} and the \texttt{qObjects} package by the authors and Schneider that is under preparation. We are receptive to adding other useful and necessary functionality to this package. Any and all updates about these additions and changes will be announced on the authors' web pages and on the RISC website. This package can be downloaded at \url{http://www.risc.jku.at/research/combinat/software/qFunctions}.

\section{Acknowledgement}

The authors would like to thank RISC and RICAM, where this work was cultivated. In particular, we want to thank Peter Paule, Carsten Schneider, Christoph Koutschan, Veronika Pillwein, Chris Jennings-Shaffer, and many others we met at the OPSFA 19 conference and elsewhere for their interest, encouragement, and support.

The authors would also like to thank the Austrian Science Fund FWF. The research of the first author is suported by the FWF SFB50-09 Project and the research of the second author is supported by the SFB50-07, SFB50-09 and SFB50-11 Projects.


\begin{thebibliography}{99}

\bibitem{Alladi_Gollnitz1}  K. Alladi, and G. E. Andrews, \textit{The dual of Gollnitz's (big) partition theorem}, Ramanujan J. 36, (2015), no. 1-2, 171-201.

\bibitem{FourPar} K. Alladi, G. E. Andrews, and A. Berkovich, \textit{A new four parameter q-series identity and its partition implications}, Inventiones Mathematicae \textbf{153} (2003), no. 2, 231-260.

\bibitem{Alladi_Gollnitz}  K. Alladi, G. E. Andrews, and B. Gordon, \textit{Generalizations and refinements of partition theorems of G\"ollnitz}, J. Reine Angew. Math., 460, (1995), 165-188.

\bibitem{Refinement} K. Alladi, G. E. Andrews, and B. Gordon, \textit{Refinements and Generalizations of Capparelli's Conjecture on Partitions}, Journal of Algebra \textbf{174} (1995), no. 2, 636-658.

\bibitem{Weighted_RR} K. Alladi, and A. Berkovich, \textit{New weighted Rogers-Ramanujan partition theorems and their implications}, Trans. Amer. Math. Soc. \textbf{354} (2002), no. 7, 2557-2577. 

\bibitem{Generalized_Schur} K. Alladi, and B. Gordon, \textit{Generalizations of Schur's partition theorem}, Manuscripta Math \textbf{79} no: 1, (1993), 113-121.


\bibitem{Andrews_Comps} G. E. Andrews, \textit{The use of computers in search of identities of the Rogers-Ramanujan type}, (1971), 377-387.

\bibitem{Theory_of_Partitions}G. E. Andrews, \textit{The theory of partitions}, Cambridge Mathematical Library, Cambridge University Press, Cambridge, 1998. Reprint of the 1976 original. MR1634067 (99c:11126)

\bibitem{Andrews_q_series} G. E. Andrews, \textit{q-series: their development and application in analysis, number theory, combinatorics, physics, and computer algebra, volume 66 of CBMS Regional Conference Series in Mathematics}, Published for the Conference Board of the Mathematical Sciences, Washington, DC; by the American Mathematical Society, Providence, RI, 1986.

\bibitem{Andrews_Baxter} G. E. Andrews, and R. J. Baxter, \textit{Lattice gas generalization of the hard hexagon model. III. q-Trinomial coefficients}, J. Statist. Phys. \textbf{47} (1987), no: 3-4, 297-330.


\bibitem{BerkovichUncu7} A. Berkovich and A. K. Uncu, \textit{Polynomial identities implying Capparelli's partition theorems}, Journal of Num. Theo. \textbf{201}, (2019), 77-107.

\bibitem{BerkovichUncu8} A. Berkovich and A. K. Uncu, \textit{Elementary polynomial identities involving $q$-trinomial coefficients}, arXiv:1810.12048.

\bibitem{BerkovichUncu9} A. Berkovich and A. K. Uncu, \textit{Refined $q$-trinomial coefficients and two infinite hierarchies of $q$-series}, arXiv:1810.06497.

\bibitem{Bringmann_Dousse} K. Bringmann, J. Dousse, J. Lovejoy, and K. Mahlburg, \textit{Overpartitions with restricted odd differences}, Electron. J. Combin. 22 (2015), no.3, paper 3.17.

\bibitem{Bringmann_JS_Mahlburg} K. Bringmann, C. Jennings-Shaffer, and K. Mahlburg, \textit{Proofs and reductions of various conjectured partition identities of Kanade and Russell}, https://doi.org/10.1515/crelle-2019-0012

\bibitem{Borodin} A. Borodin, \textit{Periodic Schur process and cylindric partitions}. Duke Math. J. \textbf{140}, no.3, (2007), 391-468.

\bibitem{Corteel} S. Corteel, and T. Welsh, \textit{The $A_2$ Rogers--Ramanujan identities revisited}, arXiv:1905.08343.

\bibitem{Dousse_revisit} J. Dousse, \textit{The method of weighted words revisited}, FPSAC 2017, Séminaire Lotharingien de Combinatoire issue 78B, paper 66.

\bibitem{Dousse_Primc} J. Dousse, \textit{On partition identities of Capparelli and Primc}, arXiv:1811.02251.

\bibitem{Dousse_Siladic} J. Dousse, \textit{Siladic's theorem: weighted words, refinement and companion}, Proc. Amer.Math. Soc., \textbf{145}, (2017),1997-2009.

\bibitem{Dousse_Unif} J. Dousse, \textit{Unification, refinement and companions of generalisations of Schur's theorem}, Analytic Number Theory, Modular Forms and q-Hypergeometric Series, Springer (2018), pp. 213-251. 
     
\bibitem{Dousse_Lovejoy} J. Dousse, and J. Lovejoy, \textit{Generalizations of Capparelli's identity}, arXiv:1702.07249.

\bibitem{Foda} O. Foda, and T. Welsh, \textit{Cylindric partitions, $W_r$ characters and the Andrews-Gordon-Bressoud identities}, Journal of Physics. A, Mathematical and theoretical, \textbf{49} (16), [164004].

\bibitem{qSeries} F. Garvan, \texttt{qseries} \textit{package}, http://www.qseries.org/fgarvan/qmaple/qseries.

\bibitem{Gasper_Rahman} G. Gasper and M. Rahman, \textit{Basic hypergeometric series}, Cambridge University Press, 2004.

\bibitem{Cylindric} I. Gessel, and C. Krattenthaler, \textit{Cylindric partitions} Trans. Amer. Math. Soc. \textbf{349} (1997), no. 2, 429-479.

\bibitem{Hirschhorn} M. D. Hirschhorn, \textit{A continued fraction}, Duke Math. J. \textbf{41}, (1974), 27-33

\bibitem{IdentityFinder} S. Kanade and M. C. Russell. \texttt{IdentityFinder} \textit{and some new identities of Rogers-Ramanujan type}. Exp. Math., \textbf{24} (4), (2015), 419-423.

\bibitem{Kanade_Russell} S. Kanade and M. Russell, \textit{Staircases to analytic sum-sides for many new integer partition identities of
Rogers--Ramanujan type}, arXiv:1803.02515.

\bibitem{Guess} M. Kauers, \texttt{Guess} \textit{A package for guessing multivariate recurrence equations}, http://www.kauers.de/software.html

\bibitem{qGeneratingFunctions} M. Kauers and C. Koutschan, \textit{A Mathematica package for q-holonomic sequences and power series}, The Ramanujan Journal, 19 (2), pp. 137-150, Springer, 2009, ISSN 1382-4090.

\bibitem{HolonomicFunctions} C. Koutschan, \textit{Advanced Applications of the Holonomic Systems Approach}, RISC, Johannes Kepler University, Linz. PhD Thesis. September 2009.

\bibitem{Kagan1} K. Kur\c{s}ung\"oz, \textit{Andrews--Gordon type series for Capparelli's and G\"ollnitz--Gordon identities}, arXiv:1807.11189.

\bibitem{Kagan2} K. Kur\c{s}ung\"oz, \textit{Andrews--Gordon type series for Kanade--Russell conjectures}, arXiv:1808.01432.

\bibitem{Kagan3} K. Kur\c{s}ung\"oz, \textit{Andrews--Gordon type Series for Schur's partition identity}, arXiv:1812.10039.

\bibitem{HYPQ} C. Krattenthaler, \texttt{HYP and HYPQ} \textit{Mathematica packages for the manipulation of binomial sums and hypergeometric series, respectively q-binomial sums and basic hypergeometric series}, J. Symbolic Comput. \textbf{20} (1995), no. 5-6, 737-744.

\bibitem{RATE} C. Krattenthaler. \texttt{RATE:} \textit{A Mathematica guessing machine}, http://mat.univie.ac.at/$\sim$kratt/rate/rate.html.

\bibitem{Nandi} D. Nandi \textit{Partition identities arising from the standard $A^{(2)}_2$-modules of level 4}. ProQuest LLC, Ann Arbor, MI, 2014. Thesis (Ph.D.) - Rutgers The State University of New Jersey - New Brunswick.

\bibitem{qZeil} P. Paule, and A. Riese,\textit{ A Mathematica q-Analogue of Zeilberger’s Algorithm Based on an Algebraically Motivated Approach to q-Hypergeometric Telescoping}, in Special Functions, q-Series and Related Topics, Fields Inst. Commun., Vol. \textbf{14}, pp. 179-210, 1997.

\bibitem{qMultiSum} A. Riese, \texttt{qMultiSum}\textit{ - A Package for Proving q-Hypergeometric Multiple Summation Identities}, Journal of Symbolic Computation 35 (2003), 349-376.

\bibitem{Mathew_Thesis} M. C. Russell, \textit{Using experimental mathematics to conjecture and prove theorems in the theory of partitions and commutative and non-commutative recurrences}, ProQuest LLC, Ann Arbor, MI, 2016. Thesis (Ph.D.) - Rutgers The State University of New Jersey   New Brunswick.

\bibitem{Sigma} C. Schneider, \textit{Symbolic Summation Assists Combinatorics}, Sem.Lothar.Combin. \textbf{56}, (2007), pp.1-36. Article B56b.

\bibitem{Sills_FinRR} A. V. Sills, \textit{Finite Rogers-Ramanujan type identities}, Electron. J. Combin. \textbf{10} (2003), Research Paper 13, 122 pp.

\bibitem{Sills_RRtools}  A. V. Sills, \texttt{RRtools}\textit{ - a Maple package for aiding the discovery and proof of finite Rogers-Ramanujan type identities}. J. Symbolic Comput. \textbf{37} (2004), no. 4, 415-448. 

\bibitem{Uncu2} A. K. Uncu, \textit{On double sum generating functions in connection with some classical partition theorems}, arXiv:1811.08261.

\bibitem{Uncu3} A. K. Uncu, \textit{A polynomial identity implying Schur's partition theorem}, arXiv:1903.01157.

\bibitem{Warnaar} S. O. Warnaar, \textit{$q$-Trinomial identities}, Jour. Math. Phys \textbf{40} (1999), 2514-2530.

\end{thebibliography}
\end{document}